# Montreal AI Ethics Institute

*An international, non-profit research institute helping humanity define its place in a world increasingly driven and characterized by algorithms*

**Website**: https://montrealethics.ai
**Newsletter**: https://aiethics.substack.com

# Making Responsible AI the Norm rather than the Exception

**A report prepared by the Montreal AI Ethics Institute for the National Security Commission on Artificial Intelligence**

*Based on insights, analysis, and original research by the Montreal AI Ethics Institute (MAIEI) staff of the* ***NSCAI Key Considerations for Responsible Development and Fielding of Artificial Intelligence*** *and supplemented by workshop contributions from the AI Ethics community convened by MAIEI on September 23, 2020.*

**Submitted on January 13, 2021**

**Primary contact for the report:**

Abhishek Gupta (abhishek@montrealethics.ai)
Founder and Principal Researcher, Montreal AI Ethics Institute
Machine Learning Engineer and CSE Responsible AI Board Member, Microsoft

# Table of Contents

# Introduction

We were particularly thrilled with the opportunity to host a public consultation session to supplement the analysis that has been done by the MAIEI staff to analyze the *Key Considerations for Responsible Development and Fielding of AI* ("Document") by the NSCAI. It is one of those initiatives that stood out in 2020 with its repeated focus on looking at the practicalities of putting Responsible AI in practice rather than just focusing on the high-level principles. There are 100+ documents on the OECD AI Policy Observatory[1] which provide overlapping principles but most still shy away from providing concrete details on what needs to be done to operationalize these ideas and providing concrete directions for future research. In that vein, the Document from the NSCAI has been an excellent resource not only for the US Government agencies but also the rest of the ecosystem.

After a thorough review of the document and integration of comments from the public consultation workshop by the staff at MAIEI along with follow-on research and conceptualization, we have come up with several recommendations that we believe will further strengthen the Document from the NSCAI and help to augment the entire ecosystem's efforts in achieving responsible AI in practice.

Our overarching mindset for this is to ***make Responsible AI the norm rather than the exception***. Keeping that in mind, we believe that the following 3 ideas will be critical for the success of the measures proposed in the Document.

1. **Alleviating friction in existing workflows**
    a. This can be achieved if we are cognizant of the friction points that we encounter in the existing engineering workflows and working with the engineering teams to unblock these will be critical in **increasing adoption**. An example of this is highlighted in some early work that MAIEI presented at ICML 2020 Challenges in Deploying and Monitoring Machine Learning Systems[2] titled Green Lighting ML: Confidentiality, Integrity, and Availability of Machine Learning Systems in Deployment[3].

2. **Empowering stakeholders and buy-in**
    a. This also requires buy-in from stakeholders (which can be achieved through highlighting the value that responsible AI brings and how it is **not a barrier to achieving organizational goals**) across the value-add chain and organizational

[1] https://oecd.ai
[2] https://icml.cc/Conferences/2020/ScheduleMultitrack?event=5738
[3] Gupta, A., & Galinkin, E. (2020). Green Lighting ML: Confidentiality, Integrity, and Availability of Machine Learning Systems in Deployment. arXiv preprint arXiv:2007.04693.

tweaks that **empower the stakeholders to take necessary action** that can move them and their teams towards building more responsible AI systems.

3. **Effective translation of abstract standards into actionable engineering practices**
    a. And finally a translation of the myriad standards, policies, principles, guidelines, and regulations into more **actionable measures** that are tied to engineering workflows will help in the integration of these notions into our everyday practices.

In addition to specific comments on the various points highlighted in the Document, which is what we begin this report with, we also make some proposals along the lines of:

1. **Learning, knowledge, and information exchange**
2. **The Three Ways of Responsible AI**
3. **Empirically-driven risk-prioritization matrix**
4. **Achieving the right level of complexity**

that we believe will help improve the efficacy of the measures outlined in the Document. All of our comments on the specific ideas in the Document are embedded in the above framework to help contextualize our recommendations.

Each of the areas follow the template of defining what we mean by that area, why it is needed, what it will look like, how to put it into practice, and how to measure the impact. This framework we believe will be essential in bolstering the focus on operationalization of the measures highlighted in the Document.

We expand on these in this report and provide more details on how it will aid in the principles, gaps, and future research directions as articulated in the Document.

Reiterating our strong endorsement of the ideas highlighted in the Document from the NSCAI, we hope that this report aids the iteration process of the Document and we are available and eager to work with the NSCAI and other US Government Agencies to ***make Responsible AI the Norm rather than the Exception***.

Abhishek Gupta

Founder and Principal Researcher, Montreal AI Ethics Institute
Machine Learning Engineer and CSE Responsible AI Board Member, Microsoft

# Some general comments

## Organizational

1. Within each of the services that have been identified as the beachhead adopters for the measures and recommendations made in the Document, **identification of the individuals who will be in-charge** is going to be important:
    a. As it will help those individuals prepare and lay down the groundwork for integrating these ideas into their organization's work
    b. It will also provide for public accountability mechanisms that will help external stakeholders get a sense of the concrete progress that is going to be made on the implementation of the ideas in the Document.
2. **Sharing the criteria for the selection of these organisations** will also be important from a public trust building perspective since this was a comment that emerged in the workshops repeatedly as a concern.
3. There would need to be a **balance between the flexibility** offered to the different US Government agencies and the degree of prescription so that there isn't too much leeway
    a. On the one hand too much prescription might hamper emergent efforts from the organizations on the ground as they field AI systems.
    b. But, having inconsistency in application of these standards will also limit the usefulness in getting a strong Responsible AI posture across all the US Government agencies.
4. The earlier that there is **shared vocabulary developed across the different US Government agencies**, the more likely it is that the efforts happening across the board can support and bolster each other.
    a. Building on the above point, when the following terms are used in the Document, trustworthy, reliable, robust, and understandable, having more accessible definitions that take into particular consideration that audience from whose perspective we are considering reliability, as an example, will help those charged with putting these ideas into practice with more clear guidance on which aspects of these ideas to prioritize.
    b. This will also help the *checks and balances* system operate more efficiently by judging when there might be a violation of the different agreed upon definitions as they relate to Responsible AI.
5. While the Document does mention training of the workforce as a critical requirement, we would like to reiterate the importance of training that accounts for human-computer interaction in a way that makes the human operators and collaborators with the automated systems competent to the point that they **don't suffer from either automation bias[4] or algorithmic aversion[5].**

   a. The token human problem is definitely something we would want to avoid in this case, as highlighted famously in the Air France 447 fatal crash where the crew was ill-prepared for a handover when the automated system disengaged[6].
   b. This also relates to the articulation of the performance standards and metrics as highlighted in the Document. The staff needs to be trained adequately to be able to interpret these metrics correctly and make informed decisions based on them.
   c. When it comes to training, having things like *Guess the Correlation*[7] would be quite useful so that people develop a better understanding of probabilities because it has been demonstrated time and again that people struggle from understanding those well.
   d. Understanding of both the fundamental concepts and the high-level concepts in terms of how the system components interact with each other is important for the efficient collaboration of human operators with their machine counterparts.
   e. In addition, training to recognize human cognitive biases so that human operators working with machines can recognize where they might be succumbing to them will help to increase the efficacy of the other measures that are going to be put in place.
   f. With these points in mind, an examination of the human-human interface as mediated by machines might also warrant investigation and research to make them more amenable to achieving the goals of Responsible AI.

6. The Document already highlights the alignment that exists between the measures and ideas proposed in the Document with the DoD principles in the Appendix, a more detailed analysis of any other principles and approaches adopted by the other US Government agencies should also be considered and shared with the public to build trust and confidence in the ability of all the agencies to work as a **harmonized system**.
7. Something that could bear **more explicit enunciation in the Document is surrounding defense and nondefense uses** specifically in relation to the thoroughness with which the Responsible AI principles will be applied.
8. Augmenting the already mentioned measures in the Document, we also recommend the **inclusion of a "bug bounty" program** that can centralize the communication of failure modes and instances so that there is a shared understanding of how and where the systems go wrong and tapping into network-wide expertise to address those failures. This can be bolstered through the use of the LKIE as elucidated in this report.
   a. This is of great importance because AI systems smoothen out small errors while making larger, more catastrophic errors much more likely and they can surface in unexpected ways making their prediction and preparation to counter their effects much harder.

[4] Parasuraman, R., & Manzey, D. H. (2010). Complacency and bias in human use of automation: An attentional integration. Human factors, 52(3), 381-410.
[5] Dietvorst, B. J., Simmons, J. P., & Massey, C. (2015). Algorithm aversion: People erroneously avoid algorithms after seeing them err. Journal of Experimental Psychology: General, 144(1), 114.
[6] http://bea.aero/docspa/2009/f-cp090601.en/pdf/f-cp090601.en.pdf
[7] http://guessthecorrelation.com/

9. In working to incorporate value judgements in the selection of the objectives of the system, the Document mentions the involvement of diverse stakeholders, but we would further add that **codifying this as a requirement in the operations of the organization** will make it an inevitable part of the everyday processes rather than having to think about it in every instance when an AI system is being procured or developed.
    a. A related concern is the problem of *faux gatekeepers* who may either claim to represent a community or have had past experiences in representing a community but don't any longer either because of a shift in their own lived experiences and/or the evolution of the needs of that community in the first place.
    b. Making sure that we are aware of these systemic failings will help to ensure that representation doesn't fall into a tokenization problem when considering the inclusion of diverse stakeholders in making determinations about the objectives of the system.
10. The capabilities and limitations of AI systems are fast-evolving and unevenly distributed across various ML subdomains and industry domains meaning that the **trend prediction capabilities of the US Government agencies will be essential** in the success of keeping the Responsible AI initiatives up-to-speed and ready to handle the emergent landscape. This can be achieved quite effectively through the proposed LKIE.
    a. Related to this is the point of keeping an eye out on adjacent fields that might have faced similar challenges in the past and can serve as good models to learn and adapt from to create solutions to challenges in the field of AI, for example bioethics.
11. While diversity is a term that will constantly evolve to incorporate new social categories as they emerge in society, having a public rubric that shares how those considerations are incorporated into the various organizations will be important to build trust but also empower public accountability.
12. From a budgetary standpoint, making mandatory line items that relate to investigation and implementation of Responsible AI will help to make them more of a norm rather than the exception as we move towards normalizing these ideas within the work of the different US Government agencies.

## Technical

1. When thinking about risk assessments as mentioned in the Document, there is also the potential to **separate out those considerations into risk and impact assessments** which will comprise different elements giving more holistic coverage. The work titled

*Examining the Black Box: Tools for Assessing Algorithmic Systems* as summarized and analyzed here[8] provides more information on this approach.

2. For all of the technical measures mentioned in the Document, an **emphasis on their viability and effectiveness** is going to be for which the LKIE as mentioned later in this report will be a critical instrument.
3. For each of the metrics and technical measures mentioned in the Document, we advocate for the **use of measurable outcomes against which success can be determined.**
    a. From a practical standpoint it will help in adherence and prevent from burn-out or desensitization because it might be perceived as a huge, unsolvable, intractable problem in trying to achieve Responsible AI.
4. In support of the technical measures proposed in the document, we would like to call out particular attention to the responsibility that the US Government agencies can take on in **creating more public benchmarks against which AI systems can be evaluated**. A great example of this is the FRVT from NIST[9].
    a. Documentation of the assumptions and limitations of the benchmarks so created will also be essential in helping those utilizing them to make sure that they will get the intended intelligence from it rather than becoming falsely confident about the system.
5. In some of the techniques that have been proposed as active measures or future areas of research, it would be useful for those utilizing the document to **have case studies where those techniques have worked and more importantly where they haven't** so that they can make appropriate determinations of which techniques to pick. In the section on LKIE, we elaborate on a few mechanisms that will enable this knowledge to be widely dispersed within the US Government agencies network.
6. In relation to the point of perhaps not using AI systems in certain scenarios, we also propose that there be **clear mechanisms for disengaging and deactivating the system when things go wrong**. This means building failsafes and backup modes that don't have to rely on continuous access to the "intelligent" elements and have graceful failures that minimize harm. This is already something that is practised in many of the warfighting systems built by the US Government agencies and borrowing from those principles to put them into practice here will aid the process of building more robust, reliable, and safe AI systems.
    a. The handovers between humans and machines, as mentioned in the Document, should be done in a way where we correctly optimize for their relevant strengths and weaknesses[10].
7. As brought up in the discussions in the workshop, participants pointed out that there is a need for having more refinement on what is meant by feedback when used in various parts of the Document, perhaps disambiguating both the social and technical connotations in their use and integration into the AI lifecycle.

[8] https://atg-abhishek.github.io/actionable-ai-ethics/examining_the_black_box.html
[9] https://www.nist.gov/programs-projects/face-recognition-vendor-test-frvt
[10] Wilder, B., Horvitz, E., & Kamar, E. (2020). Learning to Complement Humans. arXiv preprint arXiv:2005.00582.

8. As some of the techniques have been mentioned by name in the *Engineering Practices* section, the participants from the workshop also brought up the idea of having more domain and department specific guidance that would further lower the barrier for implementing these ideas in practice.

## Community

1. When taking into account due process, in line with the overarching principles of the Document to uphold American values and Rule of Law, a consideration to keep in mind is **how the second- and third-order effects of utilizing AI within a system might limit the capability to uphold the rule of law,** specifically due process, because of lack of comprehension on the part of the defendants and those who are charged with making decisions.
    a. This can happen often when there is a gap between the level of intelligibility offered by the mechanisms in the system vs. the capabilities that the individuals have in terms of their background and training to interpret the outputs from the AI systems.
    b. Some of these ideas are further elaborated in the work titled *Different Intelligibility for Different Folks*[11] that recommends tailoring the process to meet those needs as an explicit requirement.
2. A recurring concern in the workshop discussions surrounded **the lack of transparency in terms of how the documentation as mentioned in the Document would be shared publicly**, especially in light of the recommendations of having external audits, this is something that will be crucial to building a high-degree of public trust.
    a. While some information will have to be kept private for national security reasons, defaulting to keeping the information publicly accessible for higher scrutiny and evaluation will aid in more thorough evaluation of the systems under consideration as people will have the opportunity to share more insights than would otherwise be possible through a single audit.
    b. Making the classification system public that will be used to present and share the audit reports and other documentation as outlined in the Document will also be essential in building public trust.
3. Acknowledging that this document is meant to serve the US Government agencies, we would recommend **analyzing the emphasis on American Values in conjunction with that focus' impact on how it shapes the interaction with the key allies of the US.**
    a. Building on this idea, what would be the means of engagement and disengagement on the part of the US Government agencies if there is a

[11] Zhou, Y., & Danks, D. (2020, February). Different" Intelligibility" for Different Folks. In Proceedings of the AAAI/ACM Conference on AI, Ethics, and Society (pp. 194-199).

difference between the values and principles adopted by these agencies vis a vis the counterparties who might come from different ideological and political contexts.

4. As articulated in the Document, a paramount consideration should be **whether AI should be used at all.** This emerges from discussions around how sometimes simpler, explainable models have *nearly* the same performance as the more complex models.
5. An active consideration should be the degree of onus which some of these requirements will place, especially in the procurement process which might reduce market competitiveness. This is not to say that we must compromise on the safety of these systems but more so that in places where there is a high burden, **perhaps it is an opportunity to create public commons** with the tooling that will help others build on the work that the US Government agencies do in ensuring Responsible AI.
6. Community norms around what is acceptable and what isn't change over time and given the especially long shelf-life of government run systems, we advocate for building in flexibility to adopt the disallowed outcomes as mentioned in the report and having a way to keep them up-to-date over time.
7. In terms of communication of the ideas from this and other US Government agencies' efforts, the framing and language should be collaborative rather than competitive (for example, the use of the term "arms race" is commonplace in the ecosystem) which will aid in mutually beneficial collaborations and development to move us towards Responsible AI.
8. Working hand-in-hand with industry efforts will also accelerate the adoption and implementation of Responsible AI and organizations like Microsoft are already doing so by way of sharing their expertise and insights publicly.
9. Communication with the general public in terms of how AI systems are being used in the US Government agencies will also be essential to provide the right level of insights to them and break away from the trope of "killer robots" and other ill-informed narratives on how AI is used within the government.
10. Another aspect of communication that we felt is important is addressing (and dispelling) unwarranted misrepresentations of superintelligence, its emergence, and the potential threats posed by that given how distant and perhaps unlikely it is. We include this here as it was something that was brought up as a point in the workshop discussions where participants were concerned that there wasn't a mention of how those will be addressed by the US Government agencies.

## Learning, Knowledge, and Information Exchange (LKIE)

### 1. What is it?

a. The LKIE is a way to accelerate organizational knowledge which is crucial for some of the measures outlined in the Document; specifically, there is a dire need to ensure that we have a way to leverage collective insights that are gleaned through on-the-ground practice rather than letting them sit in silos across the various arms of the US Government where the measures will be put into practice.
b. Having something to the effect of *Responsible AI Champs* who are "socializers" of this knowledge will help to transfer knowledge across the different US Government agencies.
c. There are a lot of lessons that are learned from hands-on deployments and a lot of them are **transferable across domains**.
d. Creating a **repository of use cases and associated guidance** on how those challenges were addressed, say for example the benefits and shortcomings of using ART[12], SmartNoise[13], Aequitas[14], etc. in practice
e. Also, **surfacing and sharing best practices that are actionable** through this learning exchange will also help to move the idea of responsible AI from being a fuzzy notion (which is how some technical stakeholders see it) into something that is steeped in engineering fundamentals (at the same time not abandoning the social dimensions of these challenges).

## 2. Why is this needed?

### a. Reducing redundancy

i. As an example, we might have several groups across the various US Government agencies that are working on NLP applications, when it comes to mitigating bias, there are tried-and-tested techniques that one department might figure out which we can share with others so that we don't start from scratch every time we want to address bias mitigation in NLP applications.
ii. Specifically, we would want to avoid the need (as much as possible) in using techniques that are known to have failures that other teams have discovered because they are ahead in the deployment and testing of those techniques.
iii. This is inline with the recommendations made in the Document on having robust testing and monitoring infrastructure in place to ensure Responsible AI.

[12] https://github.com/Trusted-AI/adversarial-robustness-toolbox
[13] https://docs.microsoft.com/en-us/azure/machine-learning/concept-differential-privacy
[14] https://github.com/dssg/aequitas

### b. Accelerating the deployment of Responsible AI

i. Often the integration of responsible AI principles is seen as a *hindrance* to the rapid deployment of products and service because of the additional work that is required in researching and experimenting to find the techniques that might work well.
ii. Such an exchange can short-circuit the discovery process and make it easier for practitioners to include responsible AI as an integral part of their AI lifecycle.
iii. It can also help to boost the confidence that both technical and non-technical stakeholders have in the capabilities of the deployed techniques rather than having to guess the efficacy of the methods.
iv. A point that was brought up both in the workshop discussions that we hosted and in internal deliberations on the Document was being able to adequately demonstrate and evoke trust from various stakeholders which can be developed through such a process: by highlighting what techniques were considered and justifying why certain choices were made supplemented with empirical evidence, one has a higher chance of building trust.

## 3. What will it look like?

### a. RSS feed-like subscriptions

i. Linking into the US Government agencies' project management systems, having templates (somewhat akin to the Datasheets for Datasets[15] and Model Cards for Model Reporting[16]) that will be utilized to track the efficacy of the techniques that have been tried is a way to make this step more realistic. The purpose for having a standardized template is to ease the tracking of information and make it more findable and accessible.
ii. This will have tagging for the industry domain, ML sub-domain, and other metadata that will be useful for surfacing relevant items.

[15] Gebru, T., Morgenstern, J., Vecchione, B., Vaughan, J. W., Wallach, H., Daumé III, H., & Crawford, K. (2018). Datasheets for datasets. arXiv preprint arXiv:1803.09010.
[16] Mitchell, M., Wu, S., Zaldivar, A., Barnes, P., Vasserman, L., Hutchinson, B., ... & Gebru, T. (2019, January). Model cards for model reporting. In Proceedings of the conference on fairness, accountability, and transparency (pp. 220-229).

iii. Then, when we have engineering teams embarking on a new project, they can *subscribe* to this exchange with particular tags and receive up-to-date information from across the organization on the techniques that have been tried already, what worked, what didn't, and tips and tricks in effective deployment of those techniques.

## b. Engineering focused discussion fora

i. Led by practitioners, around each of the templates that have been filled out, we will have discussions diving into the ***why*** and ***how*** for ***what*** was done.
    1. We believe that this Golden Circle[17] framework for contextualizing the work is critical in effective communication, especially in a system like the US Government agencies where there might be different styles of communication and working based on functional and regional differences.

ii. These will be different from informal and abstract discussions that are had at the moment, since we would be highly use-case driven and particularly focused on the discussion of the gaps and challenges in the application of the techniques.

iii. Tagging staff who volunteer to share their expertise will also be associated with the metadata in the templates to have convenient contact points for those who are willing to guide colleagues in their own deployments.

## c. Lived-experience community consultations

i. In our work, we are no strangers to the importance of the sociotechnical considerations regarding the impacts of the work that we do. When we don't have internal staff with experiences on how the technology we develop can have consequences, we will utilize open fora to seek insights from those that have lived experiences and bring those back to assimilate into the templates and enhance the understanding of the issues over time for everyone.

ii. This will also serve the purpose of methodically bridging the chasms in our knowledge and provide clear research directions backed by evidence organizations like the JAIC, Office of Science and Technology Policy, and

[17] https://www.ted.com/talks/simon_sinek_how_great_leaders_inspire_action?language=en

the National AI Initiative Office to embark on the research and development of tools and policies.

iii. In addition, it will serve to build trust with external stakeholders that we indeed strive to keep their best interests at the core of our design, development, and deployment processes.

## 4. How will it work in practice?

### a. Integration into existing project management workflows

i. As a high-performing organization, team members have constant demands on their limited time and adding net new workflows to the processes can be onerous. To ease adoption, deep and seamless integration into their existing workflows will be crucial.

ii. This can be done with the inclusion of progressive disclosures from a UX perspective for the technical and project management staff. That is, as the project progresses, we *open* new items within the template that need to be filled out, amortizing the requirements and onus over the lifecycle of the project.

iii. In addition, this will serve the purpose of starting the whole process with micro-habits as a way to build *muscle memory* when it comes to exercising responsible AI practices on a regular basis, akin to how James Clear advocates building habits in his book *Atomic Habits*[18]*.*

### b. Part of the US Government Agencies' internal evaluations

i. Ultimately a lot boils down to creating the right incentive structures around these practices for their adoption. we propose an initial, informal inclusion of attention paid to the addressal of these concerns in the employee evaluations.

ii. This will be gradually bumped into inclusion in official criteria thus sending a clear message to all the staff that we take responsible AI as a core consideration in all the work that we do.

[18] Clear, J. (2018). Atomic habits: An easy & proven way to build good habits & break bad ones. Penguin.

5. How will we measure the impact?

   a. SRE-inspired metrics

      i. Borrowing from the well-established field of Site Reliability Engineering (SRE)[19], we can utilize metrics like Mean Time to Recovery (MTTR) among others by adapting them into how quickly we are able to detect, address, and monitor the ethics implications of our products and services.
      ii. As an example, upon pointing out that there is bias in some internal language models, how quickly are we able to redeploy pretrained models, deliver updates to the downstream customers, and share new guidance for them so that they can incorporate this in their work can be utilized as a proxy for the success of this initiative.

# The Three Ways of Responsible AI

1. What is it?

   a. Borrowing from the idea of *The Three Ways of DevOps*[20], we'd like to propose *The Three Ways of Responsible AI* centered on the following ideas:
      i. Improving the responsible posture of the overall system by helping to hold each other accountable
      ii. Providing feedback that is fast, visible, and accurate
      iii. Continual learning and implementation of Responsible AI

2. Why is this needed?

   a. Clarity on the importance of Responsible AI as a concrete endeavour

[19] Beyer, B., Jones, C., Petoff, J., & Murphy, N. R. (2016). Site Reliability Engineering: How Google Runs Production Systems. " O'Reilly Media, Inc.".
[20] Kim, G., Behr, K., & Spafford, K. (2014). The phoenix project: A novel about IT, DevOps, and helping your business win. IT Revolution.

i. From anecdotal observations, there is rising concern among practitioners within the ecosystem on the subject of Responsible AI but a lot of those notions are fragmented with varying degrees of understanding on the core issues.
ii. One of the frustrations expressed by practitioners is the lack of concrete translations of the principles into things that they can deploy in their work. This is also manifested in the hesitation to try different things when there isn't a well established framework to evaluate what is working and what isn't.
iii. The first way (*Improving the responsible posture of the overall system by helping to hold each other accountable*) which can instill the *kaizen* mindset where we strive to continuously make incremental improvements as a way to achieving a better posture for our systems.
iv. This ties in quite well with the notion of micro-habits since they are stepping stones towards exercising Responsible AI practices on a more routine basis.

### b. Actionable advice

i. The other two ways (*Providing feedback that is fast, visible, and accurate* and *Continual learning and implementation of responsible AI*) serve to operationalize the advice in a manner that produces artifacts that are reusable by others in the organization while also tying principles to concrete outcomes that are measurable in terms of product improvements.
ii. This is very much in line with the emphasis in the Document on producing tracing artifacts throughout the AI lifecycle that will help stakeholders work on the problems in a manner that is visible to others while also helping them better achieve their goals by providing them with explicit items that aid their development process.

## 3. What will it look like?

### a. Shared manifesto

i. Given that the high-level principles as we articulated them are quite simple, we can generate a shared manifesto that anyone can take a look at and provide comments on, likely through having it open in a *wiki-styled* platform.

ii. This is something that the NSCAI, JAIC, OSTP, or now the National AI Initiative Office can review periodically to allow for revisions and alignment such that it is efficacious and reflective of emerging concerns.

iii. As pointed out in the document and something was a point of discussion both at the workshop and the internal team discussions was how there are many disparate efforts and differences in how shared terms are used whereby groups communicating with each other think that they are following the same ideas but in reality they are not. The ideas highlighted in this section are thus critical in developing that shared vocabulary that will enable all these puzzle pieces to fit together nicely.

### b. Feedback templates

i. To achieve the *Second Way* we need something that helps people provide feedback to each other in a manner that meets the requirements for it to be fast, visible, and accurate to extract more value from their efforts.

ii. The templates will not only lower the barriers in terms of what needs to be included in the feedback but it will ensure that the qualities that we seek to have in that feedback are captured every time.

## 4. How will it work in practice?

### a. Colloquiums

i. Having regular colloquiums led by practitioners from different areas, especially those who are in management positions will be a way to surface how well the *rubber is meeting the road* when it comes to applying *The Three Ways* in practice.

ii. Just as we have in academic labs, over time we could emphasize the need for practitioners to attend these colloquiums as a part of *unofficial continuing education credits* in responsible AI.

iii. This notion of *unofficial continuing education credits* will also act as an incentive over time for people to engage in the material. This is quite in line with the recommendations on workforce training and recertification processes as mentioned in the Document.

b. Post-mortems

   i. Having post-mortems on projects to see how *The Three Ways* performed would be a good way to not only test the efficacy of these principles but also find places for improvement.
   ii. Such post-mortems will also be quite familiar to people across disciplines and thus lower the barrier for trying out an activity because of the familiarity with it.

c. Pre-mortems

   i. They are an opportunity to foresee and plan for some of the worst-case scenarios that might pan out with the use of AI and see what are the practices from *The Three Ways of Responsible AI* that can help to be proactive about that.
   ii. This is in line with existing practices in the IC and DoD with wargames that help to play out different scenarios and anticipate and practise various strategies to better prepare for the real-world.

d. Growing dedicated staff

   i. All of this requires time and effort and even though at the moment the team directly working on these ideas might be small, planning for growth over time as we are able to demonstrate the effectiveness of investing efforts on this front.
   ii. Horizontal scaling on the part of teams on the ground will also help to alleviate the need for an expanded team that helps to run this along with the upskilling of the staff that is on the team that can take on these functions as a part of their own roles.

## 5. How will we measure the impact?

a. Engineering feedback forms

i. Sending out engineering feedback forms to gauge to what extent fundamentals like CI/CD are being applied in different projects, we can integrate questions on Responsible AI being applied to projects within the same survey forms. This can be useful to gain an understanding of the degree of maturity of these practices across various departments and where educational and awareness efforts should be targeted.
ii. These will help to capture both the pervasiveness of application and the depth and rigor of application of these ideas in practice and try to find gaps in places that they are not being applied as such.

b. Pulse checks on internal fora

i. By applying sentiment analysis and other NLP techniques, we can also get a live pulse on whether the *Three Ways* are being actively discussed and imbibed across the organization (keeping in mind not to violate privacy of individuals who post and interact on the fora).
ii. This can supplement the data from the less frequent surveys that are sent out, this is particularly useful in the case when we have a fast-moving field such as Responsible AI which can have meaningful developments taking place within a few months, something that the surveys might be a bit slow in capturing.

c. Potential improvements in the SRE-inspired metrics

i. Though a pure causal analysis of the impact of *The Three Ways* would be hard, we could draw some conclusions from the degree of adoption of these ideas with the change in the SRE-inspired metrics.
ii. This will help to provide empirical evidence on the success of the measures being undertaken and help to make a stronger case for dedicating more resources to the deployment of Responsible AI across the organization.

# Empirically-driven risk prioritization matrix

## 1. What is it?

a. One of the places where we have seen hesitation and impediments to the implementation of Responsible AI is that it can be **overwhelming in terms of what needs to be done** (if it wasn't already done by design) and that can place additional burdens on those who are tasked with ensuring that the product gets delivered on time to customers.
b. Our proposal is to adopt an **empirically-driven prioritization matrix** that has two axes of the likelihood of occurrence of Responsible AI violations and the severity of those violations (borrowing from the world of cybersecurity)
c. In addition, having a **CVE-NVD like database to which practitioners can subscribe for updates** from whence we can dispatch learnings that might apply to problems that you are working on and solutions that might have been encountered in different parts of the organization to address those. This will evolve over time as more people add their experiences into the system and tying this to the previous point will help to make the entire process a lot more actionable.

## 2. Why is this needed?

### a. Lifecycle view

i. Addressing AI ethics concerns as we mentioned earlier can seem like a burden and potentially even orthogonal to delivering products and services on time. But, if we adopt a more pragmatic approach that looks at the entire lifecycle investment rather than just the upfront investment that needs to be made, we can arrive at a more sustainable pace for the inclusion of these practices in the everyday work of the engineers.
ii. This is already something that is fundamental to the Document and we appreciate this focus because there are many efforts ongoing at the moment that don't take this aspect into account.
iii. Ultimately, an important consideration is the consistency and persistence of application of these ideas rather than just one-off experimentation that can have small wins in the near-term but not affect the status quo too much in the long-run.

### b. Roadmap

i. The risk matrix can provide a convenient framework for technical and product managers to select those areas that pose the highest risks to the success of the product, at the same time helping to achieve the *biggest bang for the buck* in terms of the efforts that are invested by the team when it comes to deploying Responsible AI practices.

ii. They can also serve as an easy gateway into making possible the adoption of these ideas without having to overcome too much inertia.

## 3. What will it look like?

### a. Two-dimensional matrix

i. The matrix will take the form of a two-dimensional grid with the axes of severity or risks and likelihood of risks (borrowing from cybersecurity risk matrices[21]) that can help the team make fast and effective determinations on which directions to pursue.

ii. The simplicity of the matrix is also an important consideration because it lowers the barriers to adoption and makes it accessible across multiple job roles rather than requiring very specialized expertise which can decrease the potential for cross-disciplinary collaborations.

iii. Keeping in line with one of the opening remarks in this report, aiming for low friction to increase ubiquity of adoption should always be a consideration when looking to include new processes or modifying existing processes.

## 4. How will it work in practice?

### a. Gathering empirical data

i. Through the learning, knowledge, and information exchange (LKIE) that we mentioned in this report, we can associate empirical data over time with each of the use cases that are encountered by teams over time and aggregate them by industry domain and ML sub-domain.

ii. This will also help to convince the stakeholders experimenting with and deploying these measures have a degree of confidence in the

[21] https://www.csa.gov.sg/-/media/csa/documents/legislation_supplementary_references/guide_to_conducting_cybersecurity_risk_assessment_for_cii.pdf

recommendations since they will have the trust and experience of their colleagues behind it.

b. Post-mortem on the effectiveness of the matrix recommendations

i. While the generation of counterfactuals in this case is not easy, aggregating over many use-cases, we could arrive at some causal inferences on the efficaciousness of the recommendations that are utilized from the matrix evaluations, specifically, in picking which areas should be prioritized to be addressed first.

## 5. How will we measure the impact?

### a. Pulse feedback from employees

i. Just as we looked at the other places where integrating the ask for feedback on the usefulness of the measures proposed here, we can utilize engineering pulse surveys to get a sense for what works well and what doesn't.

ii. In addition, this is also an opportunity to learn about the degree of adoption across the organization and other useful metrics that come from such an exercise. This will help tailor the approaches to different departments over time perhaps and surface best practices that can help organizational units get more mileage from these measures.

### b. SRE-inspired metrics

i. Borrowing again from the other SRE-inspired metrics mentioned before, we can get a quantitative estimate of the efficacy of these measures which will be useful in justifying their adoption both internally within teams and across teams within the organization.

ii. These will provide a much-needed quantitative aspect to supplement any qualitative data that is collected on the efficacy of these measures.

iii. As one seeks to get more organizational support to implement these ideas in practice, we believe that such empirical evidence will be essential in convincing those with resource allocation powers to dedicate more firepower to the Responsible AI efforts.

# Achieving the right level of complexity

## 1. What is it?

a. Finally, we believe that there is much to be taken from Tesler's Law[22] that we could apply to the proposed measures above: achieving the right level of complexity will be essential, both from a process and backend perspective and the front-end experience of the practitioners who will be tasked with implementing the responsible AI standard in their everyday work.

## 2. Why is this needed?

### a. Exposing too much to the developers

i. The developers are the end-users in this scenario in a sense who, unless they are deeply interested in the area of Responsible AI themselves, have severe constraints on their time to have to figure out which techniques will work well and which won't.

ii. In terms of complexity, if we leave open all the choices for different kinds of software that they can use to implement Responsible AI in their products and services, we add to the burden of the end-user in having to make that choice, often with limited information on their efficacy which can lead to misguided choices.

iii. Yet, an argument can also be made that it might be premature to advocate too strongly for particular approaches when the developers on the team might have the greatest insight into how their use-case is unique compared to what has been done already at the organization.

iv. In this case, it is important to consider the right level of complexity that is exposed to the user in terms of the granularity of control that is offered to the end-user, the developer, so that they are not overburdened but at the same time have adequate agency to make the required changes as they need to better meet the needs of their project.

[22] https://en.wikipedia.org/wiki/Law_of_conservation_of_complexity

## 3. What will it look like?

### a. Progressive disclosures

i. This will involve principles of progressive disclosure as a way of assessing where we see the most effective deployments and uptake of the principles and practices as mentioned in the rest of this document.
ii. This will be sourced from the data that is gathered from the initiatives under the other pillars mentioned in the document.

### b. Feedback calibration

i. Calibrating that with the feedback gathered from different instruments like the pulse surveys will further help to tune what level of granularity works best.
ii. Keeping in mind that complexity can either be offloaded on to the user in terms of the number of choices that they have to make or abstracted away from them such that the interface appears simple to them but there is a lot of complexity hidden away which ultimately restricts the kind of control that they can exercise.

## 4. How will it work in practice?

### a. Working with design researchers

i. Given the vast expertise that is available to the US Government agencies, we can utilize the insights that design researchers have on the front of understanding what will lead to achieving the right level of complexity for the end-users, the developers of the practices developed by the Document.
ii. Some of it can borrow from typical design methodologies, but we also suspect that there is potential here for new kinds of research to be done that will boost the effectiveness of the work that we are trying to do as well with this initiative.

b. Community insights

i. Learning directly from the practitioners on the ground will also help to fine-tune this approach to better meet their needs.
ii. Additionally, those who are using it on a regular basis in their work will also have invaluable feedback that we could utilize.

### 5. How will we measure the impact?

a. SRE-inspired metrics and feedback sentiments

i. Just as is the case with the other initiatives, the impact and efficacy of these measures can be evaluated by utilizing SRE-inspired metrics and assessing the sentiments as expressed in the feedback from the developers.

## Conclusion

Through some of the general comments made at the beginning of this report that were split along refinements and steps that need to be taken to address some of the organizational, technical, and community issues along with a framework including the learning, knowledge, and information exchange (LKIE), The Three Ways of Responsible AI, empirically-driven risk-prioritization matrix, and achieving the right level of complexity for ***Making Responsible AI the Norm rather than the Exception***, we believe that the US Government agencies and Commissions like the NSCAI have a great role to play in making governments a role model in putting Responsible AI into practice.

Governments have in the past been great leaders in undertaking massive investments and efforts to bring forth technological progress that has benefited society writ large and we believe that Responsible AI is another area where the US Government can play a crucial role, leveraging its massive influence and access to resources and allies across the globe in making this a reality.

Through the recommendations made in this report, we hope that we can further enhance the excellent work that the NSCAI has done in putting together the Document and we can collectively achieve the vision of ***Making Responsible AI the Norm rather than the Exception***.